\documentclass[preprint,showpacs,preprintnumbers,amsmath,amssymb,endfloats*]{revtex4}

\usepackage{graphicx}
\usepackage{dcolumn}
\usepackage{bm}

\begin{document}

\title{Pion re-scattering in $\pi^0$ production}

\author{V. Malafaia}
 \email{malafaia@cfif.ist.utl.pt}
\author{M. T. Pe\~na}%
 \email{teresa@cfif.ist.utl.pt}
\affiliation{%
CFIF and Department of Physics, Instituto Superior T\'ecnico, 1049-001 Lisboa, Portugal
}%

\date{\today}

\begin{abstract}
This work discusses the approximations frequently used to
calculate the contribution to pion production in proton-proton scattering near threshold
from irreducible pion re-scattering.
The reference result is obtained from the quantum mechanical
reduction of the corresponding Feynman diagram to Time-Ordered Field Theory diagrams.
We conclude that the Distorted Wave Born Approximation is quite adequate at threshold energies
and even above.
We discuss the choices for the energy of the exchanged pion, both for its propagator
and for the $\pi N$ amplitude describing the re-scattering vertex. The calculation
employs a physical model for nucleons and pions --- pseudo-vector coupling for the $\pi NN$
vertex and realistic amplitudes for the $\pi N$ re-scattering and for the $NN$ transitions 
in the initial and final states. 
\end{abstract}

\maketitle

\section{Introduction}
Recently, high quality data on meson production at threshold are available,
and new measurements of relative small effects on those reactions,
as charge symmetry breaking, are accomplished. This situation motivates us to revisit technical details and approximations currently done in the theoretical calculations of the matrix elements,
independently of the use of chiral perturbation theory or meson theory with variable phenomenological content.

\begin{figure}
\includegraphics[width=8.5cm,keepaspectratio]{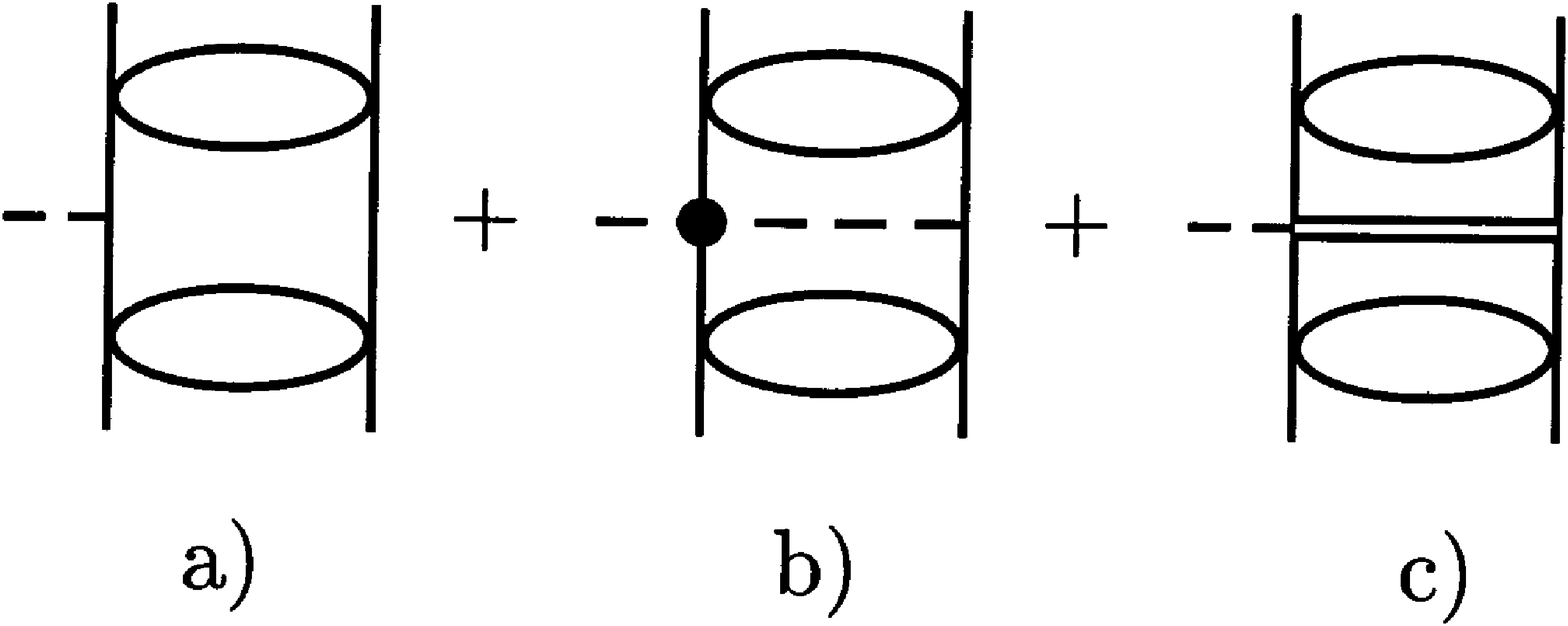}
\caption{Impulse, re-scattering and short-range processes which contribute to $\pi^0$ production.}
\label{diagrams}
\end{figure}

Interest in meson production was spurred originally by the underprediction of the cross section of the $pp \rightarrow pp\pi^{0}$ reaction. 
In Fig.\ref{diagrams} the diagrams for the impulse (Fig.\ref{diagrams}a))
and the irreducible pion re-scattering (Fig.\ref{diagrams}b)) contributions to that reaction are represented.
In the $pp \rightarrow pp \pi^{0}$ reaction,
the isospin selection rule suppresses
the  dominant low-energy isovector Weinberg-Tomosawa  $\pi$N term of 
the irreducible pion re-scattering diagram. Additionally, a negative interference between the remaining impulse  and isoscalar pion re-scattering terms\cite{Park,Cohen,Sato} may occur. Consequently,
the empirical data  for $p p \rightarrow p p \pi^0$ near
threshold is only explained by phenomenological non-pionic, thus short-range,
two-body mechanisms introduced in Ref.\cite{Riska}, or alternatively by an off-shell extrapolation of the $\pi N$ amplitude\cite{Oset,Pena2}. 
This last result, concerning the spectacular enhancement of the cross section by the off-shell behavior of the $\pi N$ 
re-scattering, was not confirmed in Ref.\cite{Hanh4} where a microscopic model for the $\pi N$ re-scattering was considered.
The high quality data for $p p \rightarrow p p \pi^0$ becomes therefore specially interesting since
it allowed to establish the importance of the short-range mechanisms represented in Fig.\ref{diagrams}c)\cite{Riska,Kolck,Hanh0,Pena}. We note that given the high momentum transfer at threshold, these
short-range mechanisms, appearing also in the short-range components of
the two and three-nucleon potentials\cite{Hanh00,Epel}, play a decisive role. Also, for the same reason, the convergence of the chiral expansion is questioned or at least calls
for a redefinition of the expansion variable, as explained in Ref.\cite{NewHanh}. 

All mechanisms described above are derived from Feynman diagrams. Nevertheless,
the evaluation of the corresponding matrix elements for the cross section proceeds through non-relativistic initial and final nucleonic wave functions, within the framework of the Distorted Wave Born Approximation (DWBA). Therefore the calculations apply a three-dimensional formulation in the loop integrals for the distortion, which is not obtained from the field-theoretical Feynman diagrams.   
Namely, the energy of the exchanged pion in the re-scattering operator (both in the $\pi N$ amplitude and in the exchanged pion propagator) has been treated approximately and under different prescriptions
in calculations performed till now.
A theoretical control of the energy for the loop integration embedding the non-relativistic reduction of the Feynman $\pi$-exchange diagram in the non-relativistic nucleon-nucleon wave functions is then still
needed, as pointed out before in Refs.\cite{Hanh1,Hanh2}.
In this paper we deal with the isoscalar re-scattering term for the $pp \rightarrow p p \pi^{0}$ reaction, near threshold.
Although in this case the
re-scattering mechanism is indeed small, the amount of its interference with the impulse term 
depends quantitatively on the calculation method.  We note, furthermore,
that the pion isoscalar
re-scattering term, which is energy dependent, increases away from threshold, 
and that for the charged pion production reactions
the isovector term is important, and depends also on the exchanged pion energy. 
Thus the knowledge gained from the application discussed here to the $pp \rightarrow pp\pi^{0}$ reaction near threshold  is useful for other applications.

Specifically, this work investigates  

i) the validity of the traditionally employed DWBA approximation. We will use as reference the result obtained from the decomposition of the Feynman diagram into 
Time-Ordered-Field-Theory (TOPT) diagrams, and realize how the last ones link naturally to an appropriate quantum-mechanical DWBA matrix element;

ii) the choices for
the exchanged pion energy which are unavoidable in the three-dimensional non-relativistic formalism underlying
DWBA, when the link described in i) is not realized. They are, namely, the {\it static} approximation, the {\it on-shell} approximation introduced first in Ref.\cite{Sato} (used in Ref.\cite{Pena} and
labeled later\cite{Hanh1} the ``$E-E'$''  approximation) and the {\it fixed kinematics} approximation\cite{Park} extensively used. 
In these three approximations
the pion energy is, respectively, taken as zero, the difference between the final and the initial
on-mass-shell energy of nucleon which emits it, and this difference calculated exactly at threshold;

iii) the numerical importance of the three-body logarithmic singularities of the exact
propagator of the exchanged pion, which are not present when the approximations mentioned in ii) are considered.

This paper generalizes the work of Refs.\cite{Hanh1,Hanh2} which considers a toy model for
scalar particles and interactions, does not include
contributions from negative-energy states, treats the nucleons as distinguishable and
therefore pion emission to proceed only from one nucleon. 
Within this toy-model, the DWBA amplitude for pion production threshold is clearly dominant over the
other contributions and, for the final-state interaction case,
only the fixed kinematics approximation (for both propagator and vertex)
leads to small deviations from the exact amplitude.
Also, in the region between the thresholds for one-pion and two-pion production,
the most used approximations for the pion propagator are found to produce the wrong energy
dependence\cite{Hanh2}.

Our calculation employs a physical model for nucleons and pions and investigates how much of the features mentioned above survive in a more realistic calculation which
uses a pseudo-vector coupling for the $\pi N N$ vertex; the $\chi$Pt $\pi N$ amplitude\cite{Park} 
and the Bonn B potential for the nucleon-nucleon interaction.
Section \ref{Sec2} makes the decomposition of the distorted irreducible pion re-scattering Feynman diagram
into the TOPT diagrams and connects the field-theoretical diagrams to the quantum-mechanical matrix elements
of DWBA; Section \ref{Sec3} presents the results and Section \ref{Sec4} a summary and conclusions.

\section{From the Feynman diagram to DWBA} \label{Sec2}
\subsection{Final-state interaction diagram}

\subsubsection{The amplitude}

\begin{figure}
\includegraphics[width=17cm,keepaspectratio]{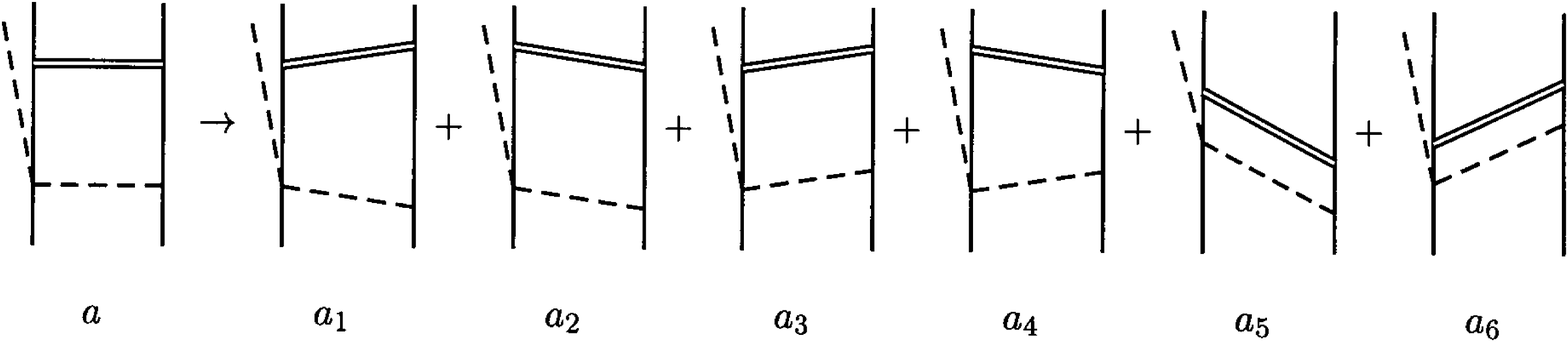}
\caption{Decomposition of the Feynman diagram in terms of six time-ordered diagrams for the final-state interaction. 
The pion(sigma) field is represented by a dashed(solid double) line. 
The nucleons are represented by solid lines. The DWBA amplitude may be identified to the first four
time-ordered diagrams ($a_1$ to $a_4$). The last two diagrams ($a_5$ to $a_6$) are usually called stretched boxes.}
\label{toptfsi}
\end{figure}

The Feynman diagram for the reaction $pp \rightarrow pp \pi^0$ where the 
$NN$ final-state interaction (FSI) proceeds
through sigma exchange is represented in part $a$ of Fig.\ref{toptfsi}.
After the nucleon negative-energy states are neglected, it
corresponds to the amplitude
\begin{eqnarray}
\mathcal{M}^{FSI} &=&  -\int \frac{d^4q'}{\left( 2 \pi \right)^4}V \left(Q'_0 \right) 
\frac{1}{Q'_0-E_2-\omega_{2}-i\varepsilon} \frac{1}{E_1-E_{\pi}+Q'_0-\omega_{1}+i\varepsilon} \label{feyfsi} \\ 
&& \frac{1}{Q'_0-E_2+F_2-\omega_{\sigma}+i \varepsilon}
\frac{1}{Q'_0-E_2+F_2+\omega_{\sigma}-i \varepsilon}\frac{1}{Q'_{0}-\omega_{\pi}+i \varepsilon}
\frac{1}{Q'_{0}+\omega_{\pi}-i \varepsilon} \nonumber,
\end{eqnarray}
where the exchanged pion has four-momentum $Q'=\left(Q'_0,\vec{q'}\right)$.
The term $V \left(Q'_0 \right)$ is a short-hand notation for the product of the $\pi NN$ vertex with the $\pi N$
re-scattering amplitude. The functions $F_{1,2}$ stand for the energy of the final nucleons,
$E_{1,2}$ for the energy of the initial nucleons and $\omega_{1,2}$, $\omega_{\pi}$, $\omega_{\sigma}$
for the on-mass-shell energy values for the intermediate nucleons, exchanged pion and sigma mesons, respectively. In Appendix A details on these functions are given.
All quantities are referred to the three-body center-of-mass frame of the $\pi N N$ final state.

In order to perform the energy integration in the variable $Q'_0$ we implemented a partial fraction decomposition
of the integrand, which isolates the poles of the pion propagator. Finally
for the integration we closed the contour in the upper-half-plane.

Because the partial fraction decomposition of the propagators in Eq.(\ref{feyfsi}) was done prior to the integration over the variable 
$Q'_0$, we have the following outcome which is independent of the choice of the contour of this integration: the only terms 
from the decomposition which do contribute to the integral correspond to the ones with only one pole, which happens to be the 
$Q'_0=\omega_\pi=\sqrt{q^{'2}+m_\pi^2}$ or the $Q'_0=-\omega_\pi=-\sqrt{q^{'2}+m_\pi^2}$ pion poles. The other terms, 
with nucleon poles and/or sigma poles, together or not with pion poles, have all these poles located on the same 
half-plane and consequently their contribution vanish.

We stress at this point that this method for the energy integration implies effectively that the $\pi N$ 
re-scattering amplitude is evaluated only for on-mass-shell pion energies. In this way, off-shell extrapolations 
which are not yet solidly constrained are avoided. Other methods may need the contribution of the off-shell amplitude in the 
integrand with the form shown in Eq.(\ref{feyfsi}). But the net result is the same, provided that all the contributions 
from all (nucleon, sigma and pion) propagator poles are considered. So far, calculations did not consider 
the pion propagator poles, since they approximate that propagator by a form free of any singularity.
Consequently, they exhibit a strong dependence on the $\pi N$ amplitude at off-mass-shell energies of the 
incoming pion.

After the $Q'_0$ integration, one obtains for Eq.(\ref{feyfsi}):
\begin{eqnarray}
&&\mathcal{M}^{FSI}_{TOPT}=-\int \frac{d^{3}q^{\prime }}{\left( 2\pi \right) ^{3}}\frac{1%
}{4\omega _{\sigma }\omega _{\pi }}\times  \label{atoptfsi} \\
&&\left[ \frac{V\left( \omega _{\pi }\right) }{\left( E_{tot}-E_{\pi
}-\omega _{1}-\omega _{2}\right) \left( E_{tot}-E_1-\omega _{2}-\omega _{\pi
}\right) \left( E_{tot}-F_1-E_\pi-\omega _{2}-\omega _{\sigma }\right) }\right.  \nonumber \\
&&\left. +\frac{V\left( \omega _{\pi }\right) }{\left( E_{tot}-E_{\pi
}-\omega _{1}-\omega _{2}\right) \left( E_{tot}-E_1-\omega _{2}-\omega _{\pi
}\right) \left( E_{tot}-F_2-E_{\pi }-\omega _{1}-\omega _{\sigma }\right) }%
\right.  \nonumber \\
&&\left. +\frac{V\left( -\omega _{\pi }\right) }{\left( E_{tot}-E_{\pi
}-\omega _{1}-\omega _{2}\right) \left( E_{tot}-E_{2}-E_{\pi }-\omega _{1}-\omega
_{\pi }\right) \left( E_{tot}-F_1-E_\pi-\omega _{2}-\omega _{\sigma }\right) }\right.  \nonumber \\
&&\left. +\frac{V\left( -\omega _{\pi }\right) }{\left( E_{tot}-E_{\pi
}-\omega _{1}-\omega _{2}\right) \left( E_{tot}-E_2-E_{\pi }-\omega _{1}-\omega
_{\pi }\right) \left( E_{tot}-F_2-E_{\pi }-\omega _{1}-\omega _{\sigma
}\right) }\right.  \nonumber \\
&&\left. +\frac{V\left( \omega _{\pi }\right) }{\left( E_{tot}-E_1-\omega
_{2}-\omega _{\pi }\right) \left( E_{tot}-F_2-E_{\pi }-\omega _{1}-\omega
_{\sigma }\right) \left( E_{tot}-E_1-F_{2}-\omega _{\pi }-\omega _{\sigma }\right) 
}\right.  \nonumber \\
&&\left. +\frac{V\left( -\omega _{\pi }\right) }{\left( E_{tot}-E_2-E_{\pi
}-\omega _{1}-\omega _{\pi }\right) \left( E_{tot}-F_1-E_\pi-\omega _{2}-\omega _{\sigma
}\right) \left( E_{tot}-E_2-F_1-E_\pi-\omega _{\pi }-\omega _{\sigma }\right) }\right] \nonumber
\end{eqnarray}
with $E_{tot}=2 E=F_1+F_2+E_\pi$ and $E = E_1=E_2$.

This equation evidences that there are six contributions to the amplitude.
These six terms, originated by the four covariant propagators of the loop,
can be interpreted as time-ordered diagrams. They are represented by diagrams $a_1$ to $a_6$ in Fig.\ref{toptfsi}.
This interpretation justifies the extra subscript label $TOPT$ for the $\mathcal{M}_{FSI}$ amplitude in Eq.(\ref{atoptfsi}).

\subsubsection{Extraction of the effective pion propagator}

From the six terms in Eq.(\ref{atoptfsi}), the first four terms (corresponding to diagrams $a_1$
to $a_4$ of Fig.\ref{toptfsi}) 
have the special feature that any cut through the intermediate state intersects only nucleon legs. Thus, they may be identified to the
traditional DWBA amplitude for the final-state distortion.
In contrast, in the last two diagrams $a_5$ and $a_6$
of Fig.\ref{toptfsi}, any cut through the intermediate state cuts not only the nucleon legs, but also two exchanged particles in
flight simultaneously. They are called the stretched boxes\cite{Hanh1}.

Because of the identification of diagrams $a_1$ to $a_4$ with DWBA, we may collect the four first terms of Eq.(\ref{atoptfsi})
and obtain what we may call the {\it exact} expression for the DWBA amplitude:
\begin{equation}
\mathcal{M}_{DWBA}=\frac{1}{2} \int \frac{d^3 q'}{\left(2 \pi \right)^3 } \left[ \tilde{V} \left( \omega_{\pi} \right)  G_{\pi}
 \right]
\frac{1}{\left(E_1+E_2-E_{\pi}-\omega_1-\omega_2 \right)} T_{NN}^{FSI}. \label{dwbafsi}
\end{equation}
Here $T$ stands for the transition-matrix of the final-state interaction. We have used both sigma exchange,
(with $m_{\sigma}=550$MeV) as in Ref.\cite{Hanh1}
\begin{equation}
V_{\sigma}^{DWBA}=\frac{1}{2 \omega_\sigma} \left[\frac{1}{\left( E_{tot}-F_1-E_\pi-\omega_2-\omega_\sigma \right)}+
\frac{1}{\left(E_{tot}-F_2-E_\pi-\omega_1-\omega_\sigma \right)} \right]
\end{equation}
which makes Eq.(\ref{dwbafsi}) coincide exactly with 
Eq.(\ref{atoptfsi}), and also the T-matrix calculated from the Bonn B potential.
In the derivation of the integrand of Eq.(\ref{dwbafsi}) the propagators for the two nucleons in the intermediate state originated fused into only one overall propagator in the non-relativistic form,
\begin{equation}
G_{NN}=\frac{1}{\left(E_1+E_2-E_{\pi}-\omega_1-\omega_2 \right)},
\end{equation}

The function $\tilde{V} \left(\omega_\pi \right)$ includes the contribution of the two pion poles
$\omega_\pi$ and $-\omega_\pi$ corresponding to two different time-ordered diagrams,
\begin{equation}
\tilde{V} = \frac{V \left(\omega_\pi \right) \left(E_1-E_\pi-\omega_1-\omega_\pi \right)+V \left(- \omega_\pi \right) \left( E_2 -
\omega_2-\omega_\pi \right)}{\omega_\pi}
\end{equation}
where the function $V \left( \omega_\pi \right)$ is the product of the $\pi N$ amplitude with
the $\pi N N$ vertex. The kinematic factors $\frac{E_1-E_\pi-\omega_1-\omega_\pi}{\omega_\pi}$ and $\frac{E_2-\omega_1-\omega_2}{\omega_\pi}$ may be interpreted as form factors.

For the realistic model used here\cite{Park} we have
\begin{equation}
V \left(\omega_\pi \right) \equiv V \left( \vec{q'},Q_\pi \right)=-\frac{1}{\left( 2 \pi \right)^3}
\frac{g_A}{f_\pi} \left(\vec{\sigma}_2 \cdot \vec{q'} \right)
\left[2 c_1-\left(c_2-\frac{g_A^2}{8M} \right)\frac{E_\pi \omega_\pi}{m_\pi^2}-c_3 \frac{E_\pi \omega_\pi -\vec{q'} \cdot \vec{q}_\pi}{m_\pi^2} \right], \label{Vpipinn}
\end{equation}
where $Q_\pi=\left(E_\pi,\vec{q}_\pi \right)$ is the four-momentum of the emitted pion.

In the derivation of Eq.(\ref{dwbafsi}) from  Eq.(\ref{atoptfsi}) the function $G_{\pi}$ for the pion propagator turns to be 
exactly 
\begin{equation}
G_{\pi}=\frac{1}{\left[\frac{\omega_1-\omega_2}{2}+\frac{E_{\pi}}{2} \right]^2- \left[\left(
E_{tot}-E-\frac{E_{\pi}}{2}\right)-\frac{\omega_1+\omega_2}{2}-\omega_{\pi} \right]^2},
\label{propagator}
\end{equation}
which gives the form of effective pion propagator appropriate for a DWBA final-state calculation and can be written as
\begin{equation}
G_{\pi}=\frac{1}{\left[{\omega_1+E_{\pi}-E+\omega_\pi}\right]\left[{E-\omega_2-\omega_\pi}\right]}.
\label{propagator2}
\end{equation} 

Thus, in what follows Eq.(\ref{dwbafsi}) is taken  as the reference result, and allows to investigate the effect of the most 
common approximations hitherto used for the pion propagator\cite{Pena2,Sato}.These approximations correspond to three different 
choices for the energy of the exchanged pion in the exchange process. In a non-relativistic framework
particles are always on-mass-shell but intermediate states can be off-energy-shell. 
Considering, however, on-energy-shell
nucleon states for the nucleons in the intermediate state of the distorted diagram, corresponds to assume that
$\omega_1=2E-E_\pi-\omega_2$.
This assumption originates the so called {\it on-shell} approximation, introduced in Ref.\cite{Sato} where the 
propagator is given by 
\begin{equation}
\begin{array}{llllll}
G^{on} & = & \frac{1}{\left( E-\omega_2 \right)^ 2-\omega_{\pi}^2 } & & &{\rm (on-shell} 
\hspace{0.1in} {\rm approximation)}.
\end{array}
\end{equation}
This has the form of the Klein-Gordon propagator, with the pion energy given by
$E-\omega_2$, i.e. the difference between the initial and
final on-mass-shell energy of  the nucleon emitting the pion.
When furthermore one takes the kinematics as the one at threshold, $E-\omega_1=E-\omega_2=E-M=\frac{E_\pi}{2}=\frac{m_\pi}{2}$, one originates the {\it fixed kinematics} approximation, where the propagator is taken as
\begin{equation}
\begin{array}{llllll}
G^{fk} & = & \frac{1}{\left( \frac{m_{\pi}}{2}\right)^ 2-\omega_{\pi}^2 } & &\hspace{1.5cm} &{\rm (fixed} \hspace{0.1in}
{\rm  kinematics}\hspace{0.1in} {\rm approximation)}.
\end{array}
\end{equation}
Naturally, the deviation between the two approximations increases with energy.
Finally, one may consider an instantaneous pion exchange process, by taking the energy of
the pion to be zero. This defines the {\it static} approximation, as
\begin{equation}
\begin{array}{llllll}
G^{st} & = & - \frac{1}{\omega_{\pi}^2 } & & &{\rm (static} \hspace{0.1in} {\rm approximation).} 
\end{array}
\label{gappfsi}
\end{equation}

It is part of the usual non-relativistic approximation for exchange Feynman diagrams.

\subsection{Initial-state interaction diagram}

\begin{figure}
\includegraphics[width=17cm,keepaspectratio]{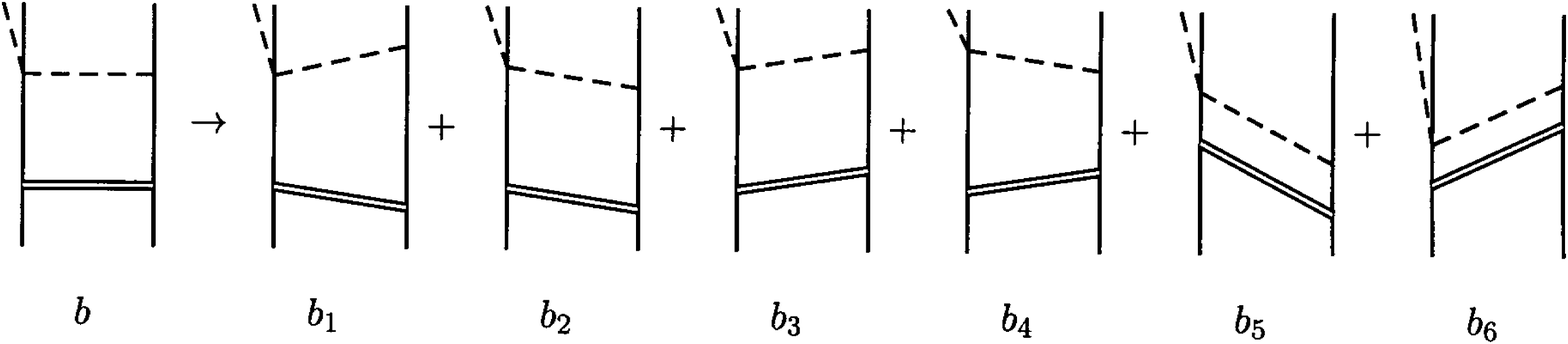}
\caption{ Decomposition of the Feynman diagram in terms of six time-ordered diagrams for the initial-state interaction.
The pion(sigma) field is represented by a dashed(solid double) line.
The nucleons are represented by solid lines. The DWBA amplitude corresponds to the first four time-ordered diagrams
($b_1$ to $b_4$) and the stretched boxes to the last two ($b_5$ to $b_6$).}
\label{toptisi}
\end{figure}

\subsubsection{The amplitude}
The corresponding  Feynman diagram (part $b$ of Fig.\ref{toptisi}), for the 
$NN$ initial-state interaction (ISI) proceeding through sigma exchange, after the nucleon negative-energy states are neglected, generates the amplitude:
\begin{eqnarray}
\mathcal{M}^{ISI} &=&  -\int \frac{d^4q'}{\left( 2 \pi \right)^4}V \left(Q'_0 \right) 
\frac{1}{Q'_0+F_2-\omega_{2}+i\varepsilon} \frac{1}{F_1+E_{\pi}-Q'_0-\omega_{1}+i\varepsilon}  \label{feyisi} \\
&& \frac{1}{Q'_0-E_2+F_2-\omega_{\sigma}+i \varepsilon}
\frac{1}{Q'_0-E_2+F_2+\omega_{\sigma}-i \varepsilon}\frac{1}{Q'_{0}-\omega_{\pi}+i \varepsilon}\frac{1}{Q'_{0}+\omega_{\pi}-i \varepsilon}
\nonumber ,
\end{eqnarray}
where we have used a notation analogous to one used for the final-state amplitude Eq.(\ref{feyfsi}).

In order to perform the integration over the exchanged pion energy $Q'_0$, as before,
a partial fraction decomposition to isolate the poles of the pion propagator was done. By closing the contour such that only the
residues of the $Q'_0=\pm\omega_\pi=\pm\sqrt{q^{'2}+m_\pi^2}$ poles contributes, one obtains the amplitude:
\begin{eqnarray}
&\mathcal{M}^{ISI}_{TOPT}&=-\int \frac{d^{3}q^{\prime }}{\left( 2\pi \right) ^{3}}\frac{1%
}{4\omega _{\sigma }\omega _{\pi }}\times  \label{atoptisi}\\
&&\left[ \frac{-V\left( -\omega _{\pi }\right) }{\left( E_{tot}-\omega
_{1}-\omega _{2}\right) \left( E_{tot}-E_1-\omega _{2}-\omega _{\sigma }\right)
\left( E_{tot}-F_1-E_\pi-\omega _{2}-\omega _{\pi }\right) }\right.  \nonumber \\
&&\left. +\frac{-V\left( \omega _{\pi }\right) }{\left( E_{tot}-\omega
_{1}-\omega _{2}\right) \left( E_{tot}-E_1-\omega _{2}-\omega _{\sigma }\right)
\left( E_{tot }-F_{2}-\omega _{1}-\omega _{\pi }\right) }\right.  \nonumber \\
&&\left. +\frac{-V\left( -\omega _{\pi }\right) }{\left( E_{tot}-\omega
_{1}-\omega _{2}\right) \left( E_{tot}-E_2-\omega _{1}-\omega _{\sigma }\right)
\left( E_{tot}-F_1-E_\pi-\omega _{2}-\omega _{\pi }\right) }\right.   \nonumber \\
&&\left. +\frac{-V\left( \omega _{\pi }\right) }{\left( E_{tot}-\omega
_{1}-\omega _{2}\right) \left( E_{tot}-E_2-\omega _{1}-\omega _{\sigma }\right)
\left( E_{tot}-F_{2}-\omega _{1}-\omega _{\pi }\right) }\right.  \nonumber \\
&&\left. +\frac{-V\left( \omega _{\pi }\right) }{\left( E_{tot}-E_1-\omega
_{2}-\omega _{\sigma }\right) \left( E_{tot}-E_1-F_{2}-\omega _{\pi }-\omega
_{\sigma }\right) \left( E_{tot }-F_{2}-\omega _{1}-\omega _{\pi }\right) }%
\right.   \nonumber \\
&&\left. +\frac{-V\left( -\omega _{\pi }\right) }{\left( E_{tot}-E_2-\omega
_{1}-\omega _{\sigma }\right) \left( E_{tot}-F_1-E_\pi-\omega _{2}-\omega _{\pi }\right)
\left( E_{tot}-E_{2}-E_{\pi }-F_{1}-\omega _{\pi }-\omega _{\sigma }\right) }\right] \nonumber 
\end{eqnarray}

The six terms in Eq.(\ref{atoptisi}) are interpreted as contributions from time-ordered diagrams represented in
Fig.\ref{toptisi}, $b_1$ to $b_6$.
This interpretation justifies the extra subscript label $TOPT$ for the $\mathcal{M}^{ISI}$ amplitude.
Although Eq.(\ref{atoptfsi}) and Eq.(\ref{atoptisi}) are formally alike, for the initial state an
extra pole is present (besides that from the nucleons propagator),
since it is energetically allowed for the exchanged pion to be on-mass-shell.
All the singularities are decisive for the real and imaginary parts of Eqs.(\ref{atoptfsi}) and (\ref{atoptisi}) and  were handled 
numerically through subtraction methods (Appendix B).  

\subsubsection{Extraction of the effective pion propagator}

Analogously to the final-state case, the first four terms in Eq.(\ref{atoptisi}) (diagrams $b_1$ to $b_4$ of Fig.\ref{toptisi}, 
where any cut of the intermediate state intersects only nucleon lines) are identified with the DWBA
amplitude for the initial-state distortion, and the last two (diagrams $b_5$ and $b_6$ of Fig.\ref{toptisi}, 
with two exchanged particles in flight in any cut of the
intermediate state) to the stretched boxes.
In other words, the decomposition obtained in Eq.(\ref{atoptisi}) allows 
to write the exact or reference expression for the DWBA amplitude for the initial-state distortion, by collecting the four first terms,
which have intermediate states without exchanged particles in flight, i.e.,
\begin{equation}
\mathcal{M}_{DWBA}=-\frac{1}{2} \int \frac{d^3 q'}{\left(2 \pi \right)^3 } \left[ \tilde{V} \left( \omega_{\pi} \right)  G_{\pi} \right]
\frac{1}{\left(E_1+E_2-\omega_1-\omega_2 \right)} T_{NN}^{ISI}, \label{dwbaisi}
\end{equation}
where
\begin{equation}
\tilde{V} \left(\omega_\pi \right) = \frac{V \left(\omega_\pi \right)\left(E_\pi+F_1-\omega_1-\omega_\pi \right)
+V \left(-\omega_\pi \right) \left( F_2-\omega_2-\omega_\pi \right) }{\omega_\pi}
\end{equation} 
and again $V \left(\omega_\pi \right)$ stands for the product of the $\pi N$ amplitude with the $\pi N N$ vertex (Eq.(\ref{Vpipinn})). 

In Eq.(\ref{dwbaisi}) we made the replacement of the exchanged particle potential by the $NN$ transition-matrix and the two-nucleon propagator 
\begin{equation}
G_{NN}=\frac{1}{\left(E_1+E_2-\omega_1-\omega_2 \right)}
\end{equation} 
is the non-relativistic global $NN$ propagator.
As we did for the final-state interaction, we extracted from Eq.(\ref{dwbaisi}) the 
effective pion propagator to be included in a DWBA-type calculation.
It reads for the initial-state distortion,
\begin{equation}
G_{\pi}=\frac{1}{\left[\frac{\omega_2-\omega_1}{2}+\frac{F_1-F_2}{2}+\frac{E_{\pi}}{2} \right]^2- \left[\left(
\frac{E_{\pi}}{2}-\frac{\omega_1+\omega_2}{2}+\frac{F_1+F_2}{2}\right)-\omega_{\pi} \right]^2}.
\end{equation}
This is the reference form for the pion propagator for the ISI case to be compared with the mostly used approximations in the literature, already discussed in detail for the FSI case:
\begin{equation}
\begin{array}{lllll}
G^{fk} & = & \frac{1}{\left( \frac{m_{\pi}}{2}\right)^ 2-\omega_{\pi}^2 }&  & {\rm (fixed} \hspace{0.1in}{\rm  kinematics}\hspace{0.1in} {\rm approximation)}\\
G^{on} & = & \frac{1}{\left(\omega_2-F_2 \right)^ 2-\omega_{\pi}^2 }  & &{\rm (on-shell}
\hspace{0.1in} {\rm approximation)} \\
G^{st} & = & - \frac{1}{\omega_{\pi}^2 } & & {\rm (static} \hspace{0.1in} {\rm approximation).}
\end{array}
\end{equation}
We tested these approximations numerically,
as we did for the corresponding ones in Eq.(\ref{gappfsi}) referring to the final state.
In the next section we present the results obtained.

\section{Results} \label{Sec3}

\subsection{Stretched Boxes vs. DWBA}

\begin{figure}
\centering
\includegraphics[width=8.5cm,keepaspectratio]{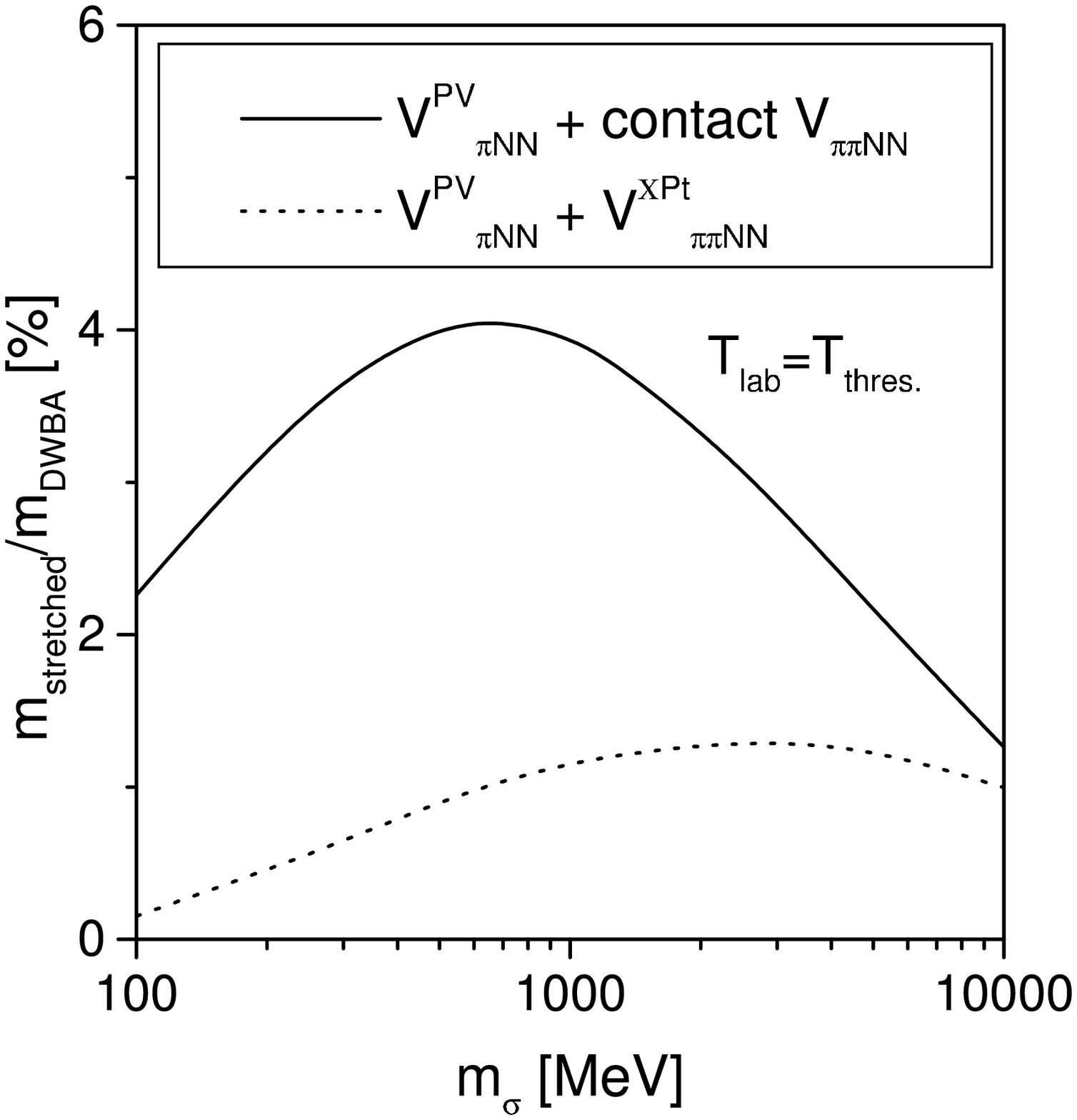}
\caption{Importance of the stretched boxes compared to the DWBA FSI amplitude $\mathcal{M}^{FSI}_{TOPT}$, 
as a function of the mass of the scalar particle for the final $NN$ interaction. 
The $\pi N $ amplitude is
contact re-scattering vertex (solid line) and the $\chi$Pt  (doted line).
The energy is taken at the pion production threshold. Absolute values of the amplitudes are considered.}
\label{mfsi}
\end{figure}

In all calculations the $NN$ and $(NN) \pi$ channels considered 
refer to the transition $^3P_0 \rightarrow \left(^1S_0 \right)s_0$, which is expected to be dominant. 
We found that the DWBA amplitude is clearly dominant over the stretched boxes in the realistic model considered, as the dotted line 
in Fig.\ref{mfsi} documents for the FSI case. 
It is interesting to compare  this result with the one
of Ref.\cite{Hanh1}. As shown in Fig.\ref{mfsi}, the stretched boxes amplitude is less than $1 \% 
$,
of the total amplitude 
and therefore is about 6 times more suppressed than in the dynamics of the toy model used in that reference.
Replacing the $\pi N$ amplitude from $\chi$Pt
by a simple contact amplitude the stretched boxes amplitude makes those boxes slightly more important,
still they do not exceed $4 \% 
$ of the DWBA amplitude (solid line in Fig.\ref{mfsi}).

\begin{figure}
\centering
\includegraphics[width=17cm,keepaspectratio]{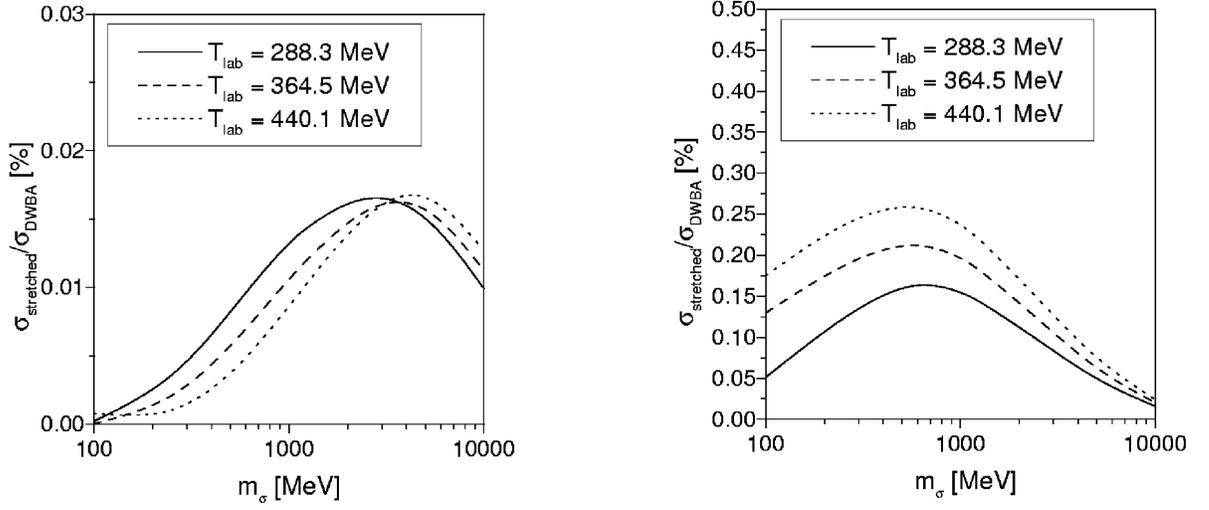}
\caption{Importance of the stretched boxes compared to the DWBA (FSI) for the cross section as a function of the mass 
of the scalar particle for the final $NN$ interaction. The $\pi N$ amplitude is the $\chi$Pt  
amplitude (left) and a contact re-scattering vertex (right).}
\label{sfsi}
\end{figure}

In terms of the cross section, 
the weight of the stretched boxes relatively to DWBA is even smaller, of the order of $0.02 \%$ at most.  
This is seen in Fig.\ref{sfsi} where we compare, for three different values of the energy, the cross section  
obtained with only the DWBA contribution 
($\sigma_{DWBA}$), with the ones obtained with only the stretched boxes terms ($\sigma_{stretched}$).
In both cases considered, $V_{\pi N N}^{PV} +V_{\pi \pi N N}^{\chi Pt} $ and
$V_{\pi N N}^{PV}$ + contact, 
respectively left and right panel in Fig.\ref{sfsi}, 
the cross section with the DWBA terms is clearly dominant, and in a more pronounced way when compared to
the less realistic case of Ref.\cite{Hanh1}, not only at threshold but even for higher energies as $440$ MeV.

However, we note that the ratios and energy dependence of the amplitudes and of the cross sections are significantly influenced by the $\pi N$ amplitude used in 
the calculation. The stretched boxes are seen to be amplified by the contact $\pi N$ re-scattering
amplitude, due to an interplay between the $\pi N$ and the $NN$ amplitudes. Relatively to the more realistic $\chi Pt$ amplitude, the contact $\pi N$ amplitude gives a larger weight to the low-momentum transfer. A realistic $\pi N$ amplitude satisfies chiral symmetry. This implies cutting small momenta and giving more weight to the region of large momentum transfer, which however in turn is cut by the nucleonic interactions. The difference between the two $\pi N$ amplitudes is clearly seen 
by comparing the behavior of each curve on the left panel of Fig.\ref{sfsi} with the corresponding curves on the right panel, for small values of the $NN$ interaction cut-off.  

\begin{figure}
\centering
\includegraphics[width=8.5cm,keepaspectratio]{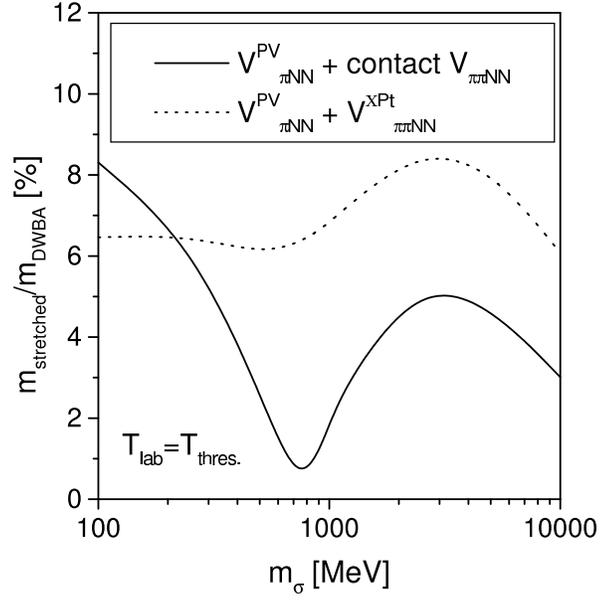}
\caption{The same as Fig.\ref{mfsi} but for the DWBA ISI amplitude $\mathcal{M}^{ISI}_{TOPT}$.
}
\label{misi}
\end{figure}

\begin{figure}
\centering
\includegraphics[width=17cm,keepaspectratio]{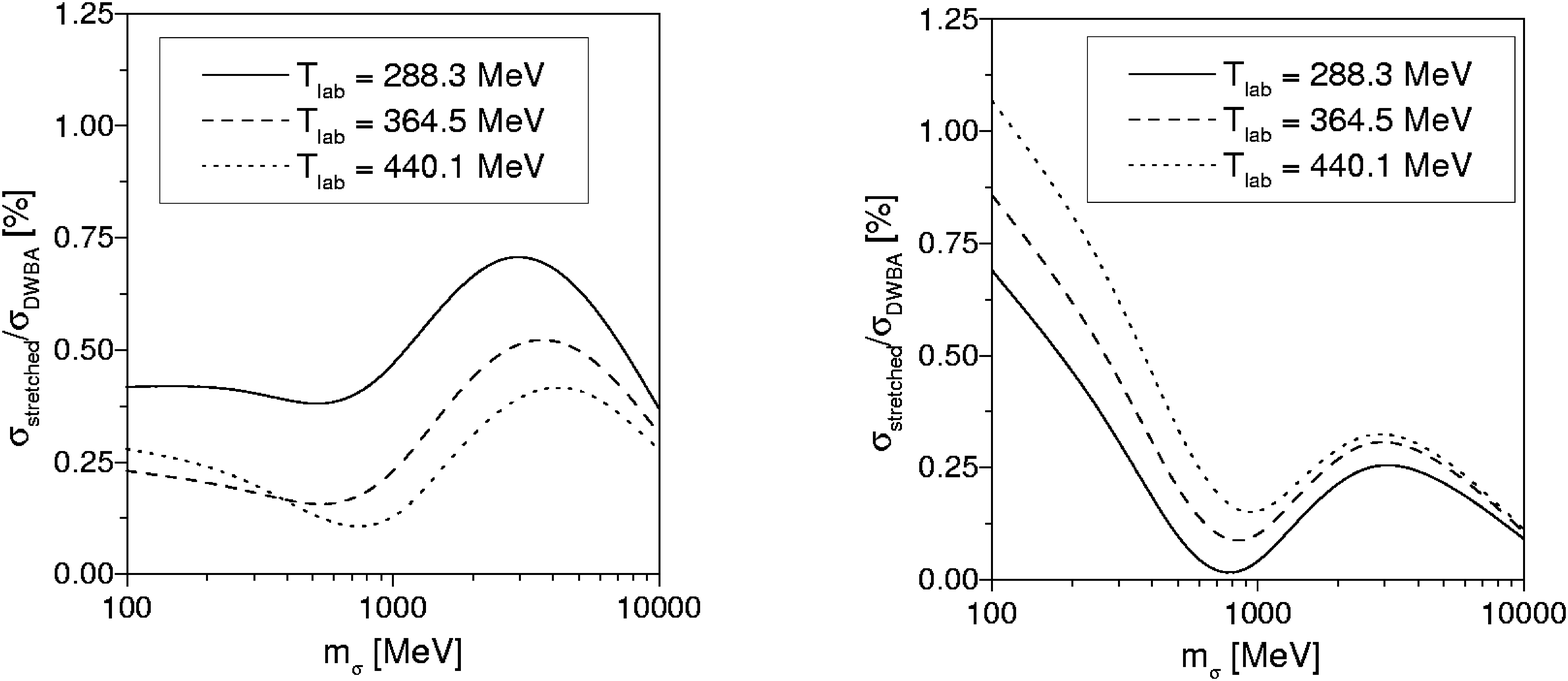}
\caption{The same as Fig.\ref{sfsi} but for the ISI case.}
\label{sisi}
\end{figure}

In the case of the initial-state amplitude, the stretched
boxes amplitude is also much smaller than the DWBA amplitude
for the two cases shown in Fig.\ref{misi}.
The cross sections obtained with only the stretched boxes terms were found to be less than 
$1.2\%
$ of the DWBA cross sections,
even for laboratory energies up to $T_{lab} \sim 440$MeV (see Fig.\ref{sisi}).

The results presented justify the  DWBA treatment for pion production.

\subsection{Approximations for the energy of the exchanged pion}

\begin{figure}
\centering
\includegraphics[width=7.6cm,keepaspectratio]{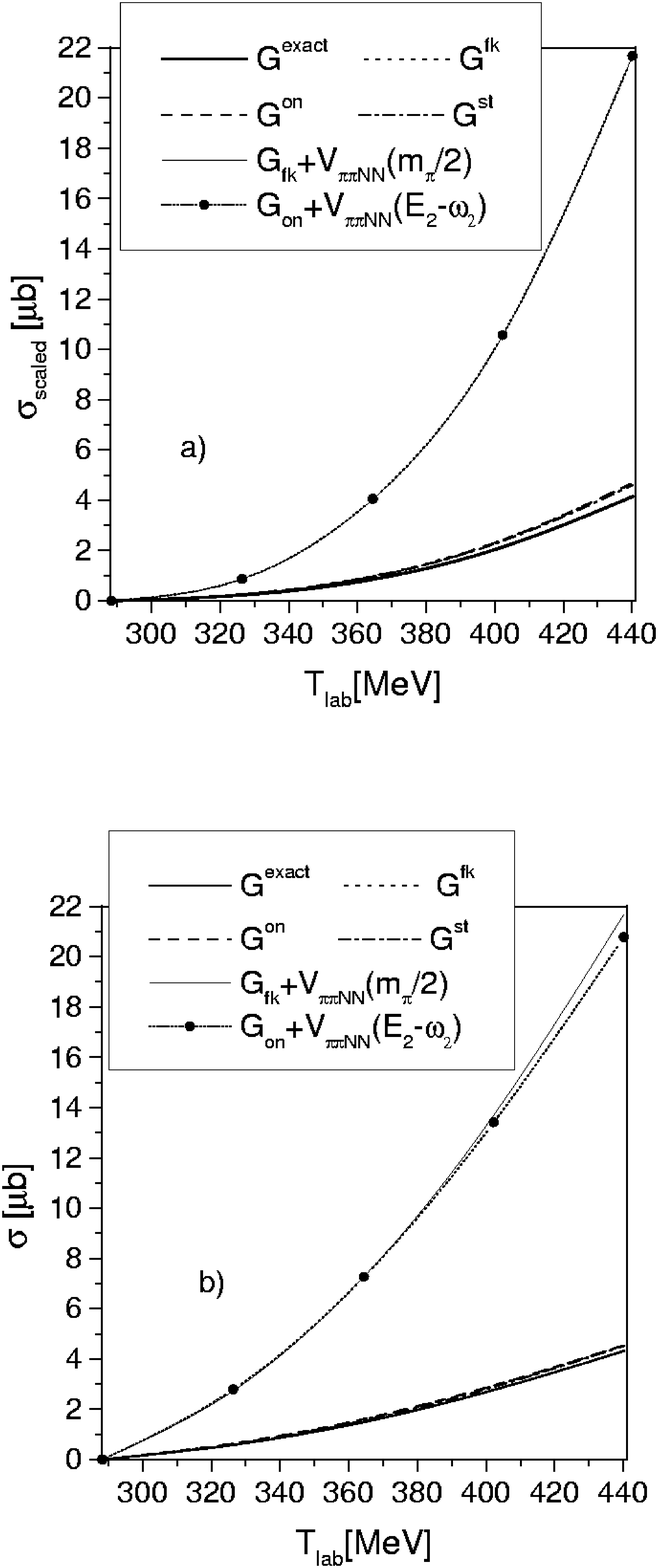}
\caption{Effect of the choices for the energy of the exchanged pion when the $\chi$Pt $\pi N$ re-scattering amplitude is considered, as a function of the laboratory kinetic energy. The nucleon-nucleon interaction is sigma exchange (panel a)) and 
the Bonn B potential (panel b)). The cross section curves shown correspond to FSI alone. 
The heavy full line is the reference calculation (Eq.(\ref{dwbafsi})). The dashed-dotted line is the {\it static} approximation. 
The dotted line is the {\it fixed 
kinematics} prescription and the dashed line is the {\it on-shell} prescription. The light full line and the light dotted line with 
bullets correspond respectively 
to the last two prescriptions, taken not only for the propagator but also for the $\pi N$ re-scattering  vertex.}
\label{gfsi}
\end{figure}

Considering the final $NN$ state-interaction case, within the model with scalar
particles and interactions solved in  Refs.\cite{Hanh1,Hanh2} all the approximations for the
pion energy taken at the $\pi N$ re-scattering amplitude (absorbed in the function $V$ of Eq.(\ref{dwbafsi}))
overestimate the cross section.
We found that these conclusions still hold in the more realistic case of a pseudo-vector pion-nucleon coupling with
the $\chi$Pt $\pi N$ re-scattering amplitude.  This is shown  in Fig.\ref{gfsi}. We remark however that
realistic $\pi N$ and $NN$ FSI distortion makes the deviation between 
exact and approximate results to be very small.
Since the Bonn potential (as all OBE potentials) corresponds
to a non-relativistic reduction where a specific method was applied to make it energy independent, 
we tested the effect on the results of an interaction in the final state corresponding
to an energy-dependent sigma exchange. The results are on the panel a) of Fig.\ref{gfsi}.
The conclusions are about the same when we describe the nucleon-nucleon interaction by the (energy
independent) Bonn B potential scattering T-matrix, instead of  the 
sigma exchange.

In particular the results for the reference calculation (heavy full line), the {\it static} approximation (dashed-dotted line), and 
the {\it fixed 
kinematics} prescription (dotted line) on Fig.\ref{gfsi} practically coincide. This means that
the choice of energy for the pion propagator is not very decisive for the FSI amplitude, confirming that the usual low-energy static or
instantaneous approximation for exchange diagrams is justified in that case.

We note however from Fig.\ref{gfsi} that the cross section is dramatically enhanced if the ``ad-hoc" 
fixed kinematics energy is considered at the same time for the re-scattering
amplitude and the pion propagator. The results show indeed that the fixed kinematics 
(light full line in Fig.\ref{gfsi}) and in particular the on-shell prescription (light dotted line with bullets), when 
taken both for the re-scattering vertex energy and pion propagator, are nor adequate above threshold, even at energies close to it, 
and they overestimate the cross section by a large factor. This finding is consistent with
the results of Ref.\cite{Sato} where the on-shell approximation for the energy at the 
$\pi N$ amplitude and the pion propagator was introduced for the first time. 

\begin{figure}
\centering
\includegraphics[width=7.6cm,keepaspectratio]{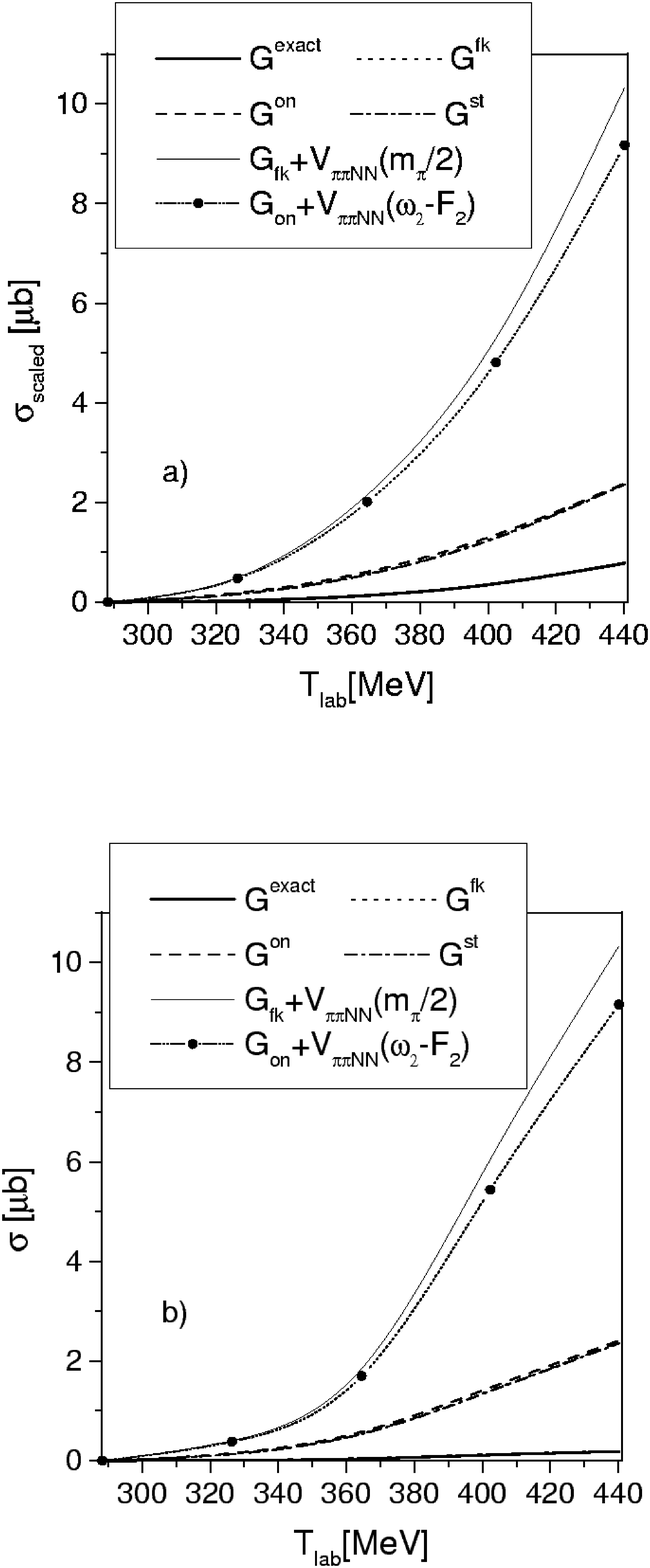}
\caption{The same as Fig.\ref{gfsi} but for the ISI amplitude.}
\label{gisi}
\end{figure}

As for the initial-state distortion, our calculations checked the expected result
that the cross section for the initial-state interaction is much smaller
than the cross section for the final-state interaction.
Still, regarding the approximations for the energy of the exchanged pion, 
the cross sections with the various pion propagator prescriptions defined in the previous section
follow the trend observed for the final state amplitude. In particular,  
when the $\chi$Pt $\pi N$
re-scattering amplitude is considered all the approximations for
the propagator in the ISI amplitude overestimate the exact cross section, given by the heavy full line
on Fig.\ref{gisi}.  This happens
for a  nucleon-nucleon interaction given by the (energy independent) Bonn B potential (panel b) of Fig.\ref{gisi}), or by the energy dependent sigma exchange (panel a) of Fig.\ref{gisi}).

When we compare the results for the initial state interaction with the results obtained for the final state interaction, we recognize
however that the deviation between the approximate results and the exact one is much more 
significant for the ISI amplitude.
The on-shell $G^{on}$ approximation (and also the fixed kinematics approximation) gives a larger cross section (panel b) in 
Fig.\ref{gisi}). Its effect is much more pronounced than it was for the final-state distortion. This difference can be understood since the initial state interaction induces high off-shell energies 
in the intermediate nucleons. This enlarges the gap between the $G^{on}$ or $G^{fk}$ and  the $G^{exact}$
calculations, compared to what happens for the FSI case, where the nucleons
emit the outcoming pion before their interaction happens.

\begin{figure}
\centering
\includegraphics[width=8.5cm,keepaspectratio]{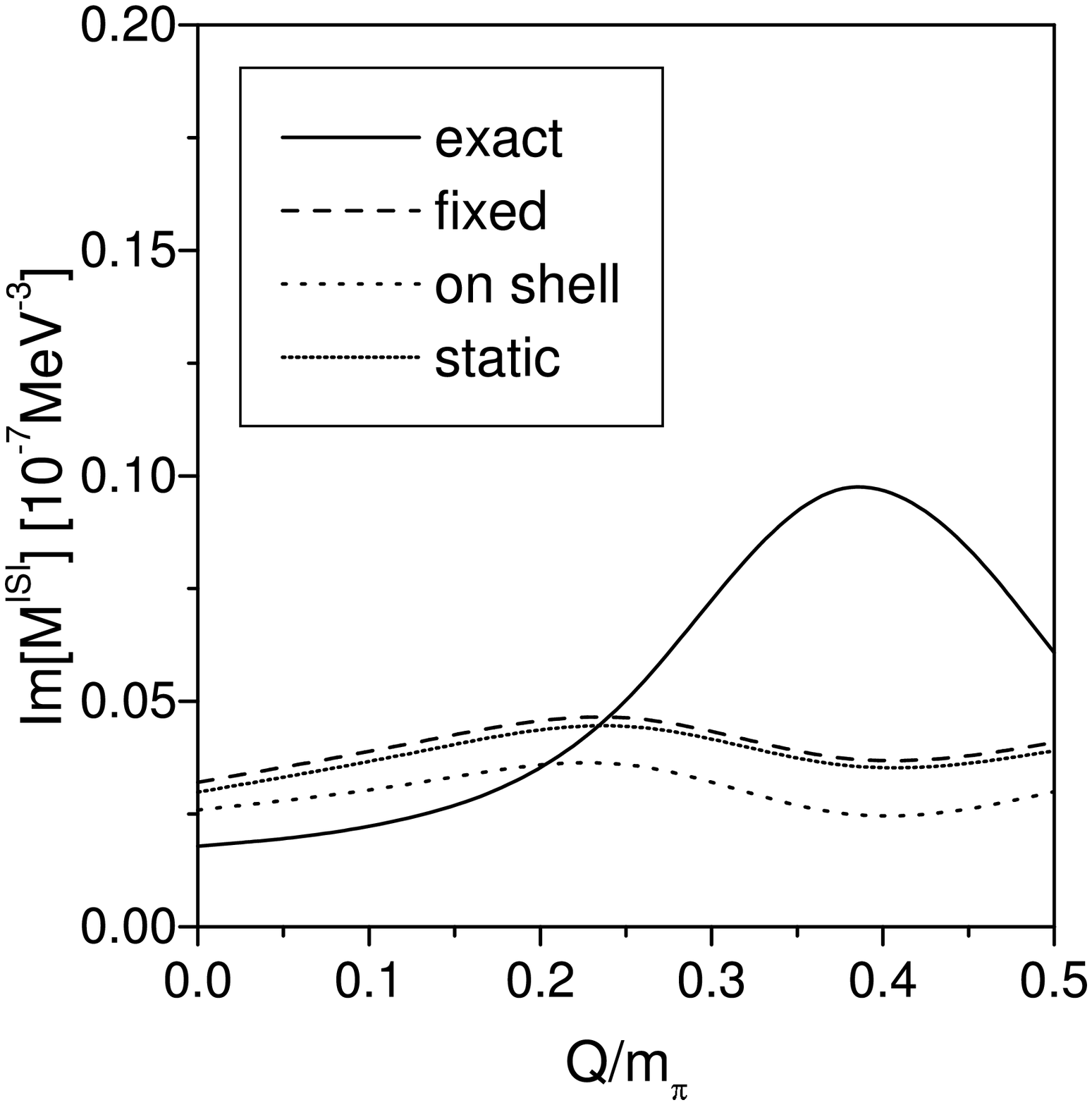}
\caption{Imaginary part of the ISI amplitude $\mathcal{M}_{DWBA}$ as a function
of the dimensionless pion momentum.
The exact result (full line) and {\it fixed kinematics} (dashed line), {\it on-shell}
(short-dashed line) and {\it static} (dotted line) approximations are shown.}
\label{imagisi}
\end{figure}

\begin{figure}
\centering
\includegraphics[width=8.5cm,keepaspectratio]{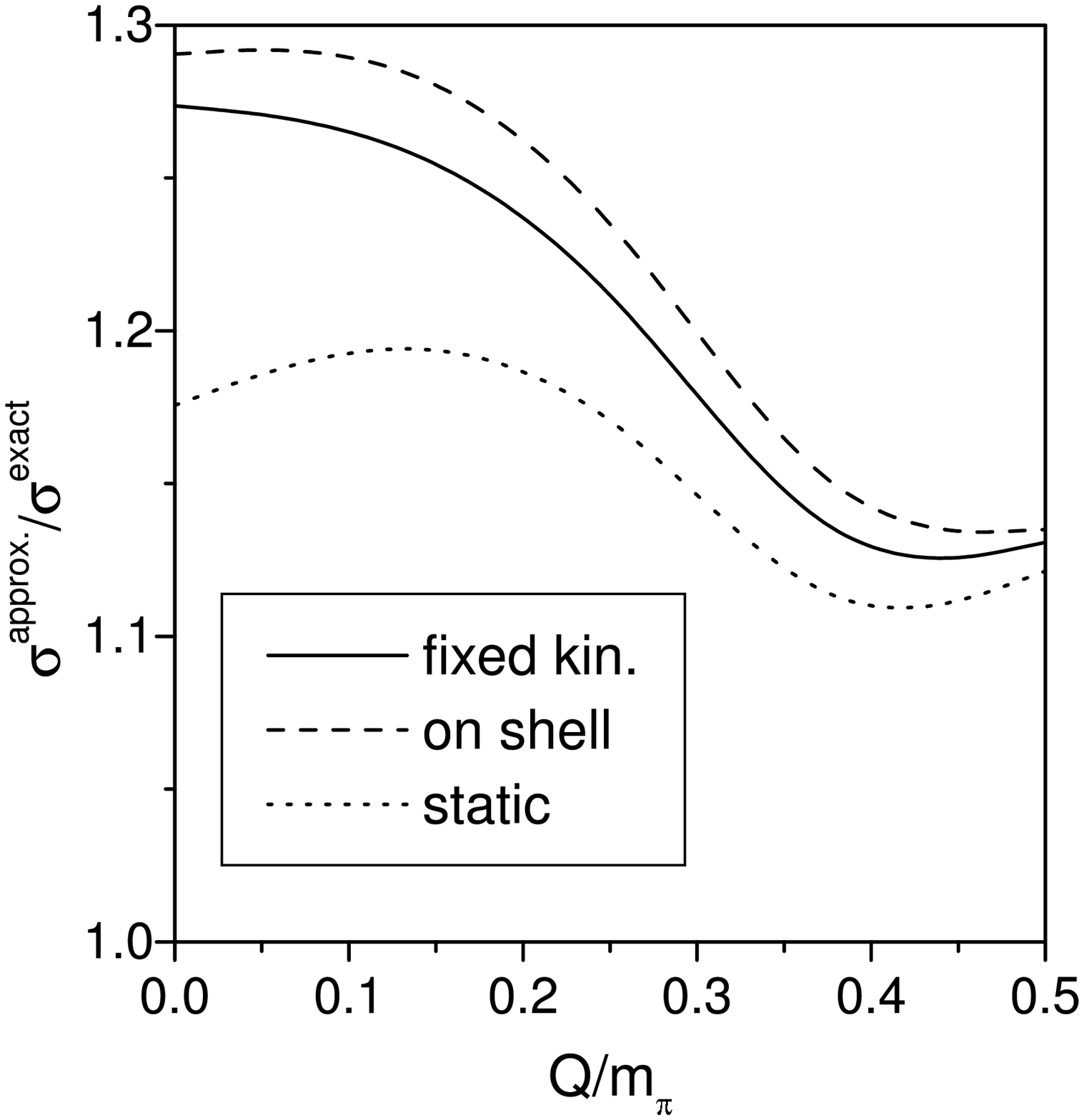}
\caption{Ratio between total approximated and exact cross section as a function of the dimensionless pion momentum. The {\it fixed kinematics} (full line), {\it on-shell}
(dashed line) and {\it static} (dotted line) cases are shown.} 
\label{totalsig}
\end{figure}

Also, what is behind the two approximations $G^{on}$ or $G^{fk}$ being worse representations of the exact amplitude for the ISI case than for the FSI case,  is the physics related to the logarithmic singularities of the exact pion propagator. These singularities are present only for
the ISI amplitude, but are absent in the case of the approximate forms for that propagator. Their
effect is visualized in Fig.\ref{imagisi} which represents the imaginary part of that amplitude.
On Fig.\ref{totalsig} we represent the ratio of the total cross sections (from the FSI, ISI and
their interference amplitudes)  for three approximations considered by the total exact cross section. The dominance of the FSI amplitude for the cross section makes the fixed-kinematics and the on-shell approximation very close at threshold, since for that amplitude the two
approximations coincide, very near threshold. The conclusion is that the total cross section can be affected
as much as $30 \%$ by the energy choices for the exchanged pion, and that the effect is largest close
to threshold.

\section{Summary and Conclusions}\label{Sec4}
We have generalized the work done in Refs.\cite{Hanh1,Hanh2} on the study of the approximations 
of the quantum-mechanical calculation of irreducible pion re-scattering on the
reaction $pp \rightarrow pp \pi^0$. We considered a physical model for nucleons and pions, combining
a pseudo-vector coupling for the $\pi N N$ vertex, the $\chi$Pt $\pi N$ re-scattering amplitude, 
and a realistic potential for the $NN$ interaction in the final and initial state. 
We found that the effect of the usual choices for the pion energy in the $\pi N$ re-scattering vertex, which is not fixed by a 
non-relativistic formalism, and thus has to be derived from the four-dimensional relativistic approach,
can be significant.

Both for the final- and initial-state interaction, our results show that
\begin{itemize}
\item[i)] the DWBA formalism is quite adequate at threshold and even at higher energies, since this part of the full amplitude is clearly seen to be
dominant over the stretched boxes. 
This result confirms quantitatively the usual DWBA treatment for pion production. This is independent of the model for the $\pi N$ 
re-scattering amplitude. Nevertheless, relatively to less realistic models, a  chirally constrained amplitude reinforces even more
the relative importance of the DWBA amplitude in the total amplitude;
\item[ii)] the choice for the energy in the pion propagator is not very decisive in the
amplitude for the final state distortion. However, the
amplitude for the distortion of the initial state shows an important sensitivity to this choice.
The approximate cross section can deviate as much as 30$\%
$ 
from the exact result.  
\item[iii)] the choice for the energy of the exchanged pion in  the re-scattering vertex,
as prescribed from an extraction from the Feynman diagram, is crucial for the cross section strength. 
The fixed kinematics and the on-shell prescriptions for the re-scattering vertex overestimate
overwhelmingly the exact amplitude.
\end{itemize}

We conclude that for evaluation of the irreducible pion re-scattering, one has to use
the reference amplitudes given by  Eqs.(\ref{dwbafsi}) and (\ref{dwbaisi}). The latter were obtained
as non-relativistic reductions of the corresponding Feynman diagrams for the final- and initial-state interactions, respectively. In the future, the reference amplitudes corresponding to
the resonance contributions for the charged $\pi$ or the $\eta$ meson production reactions,
should be analyzed in the same way.

\begin{acknowledgments}
The work of M. T. P. was supported by FCT under the grant CERN/FIS/43709/2001. 
The work of V. M. was supported by FCT under the grant SFRD/BD/4876/2001.
\end{acknowledgments}

\appendix
\section{Kinematics}
For the final state $NN$ interaction, the expressions for the
on-shell energies of the particles are:
\begin{equation}
\begin{array}{lll}
\omega _{\pi }=\sqrt{m_{\pi }^{2}+\left| -\vec{p}+\frac{\overrightarrow{%
q_{\pi }}}{2}+\overrightarrow{q_{k}}\right| ^{2}} & \qquad  & \text{%
exchanged }\pi  \\ 
\omega _{1,2}=\sqrt{M^2+\left| \overrightarrow{q_{k}}\mp \frac{%
\overrightarrow{q_{\pi }}}{2}\right| ^{2}} &  & \text{intermediate nucleons}
\\ 
\omega _{\sigma }=\sqrt{m_{\sigma }^{2}+\left| \overrightarrow{q_{k}}-%
\overrightarrow{q_{u}}\right| ^{2}} &  & \sigma \text{ exchanged} \\ 
F_{1,2}=M+\frac{1}{2M}\left| \overrightarrow{q_{u}}\mp \frac{%
\overrightarrow{q_{\pi }}}{2}\right| ^{2} &  & \text{final nucleons}
\end{array}
\end{equation}
For the initial $NN$ interaction, the corresponding expressions are
\begin{equation}
\begin{array}{lll}
\omega _{\pi }=\sqrt{m_{\pi }^{2}+\left| \overrightarrow{q_{u}}+\frac{%
\overrightarrow{q_{\pi }}}{2}-\overrightarrow{q_{k}}\right| ^{2}} & \qquad 
& \text{exchanged }\pi  \\ 
\omega _{1,2}=\sqrt{M^2+\left| \overrightarrow{q_k} \right|^2} &  & 
\text{intermediate nucleons} \\ 
\omega _{\sigma }=\sqrt{m_{\sigma }^{2}+\left| \overrightarrow{p}-%
\overrightarrow{q_{k}}\right| ^{2}} &  & \sigma \text{ exchanged} \\ 
F_{1,2}=M+\frac{1}{2M}\left| \overrightarrow{q_{u}}\mp \frac{\overrightarrow{%
q_{\pi }}}{2}\right| ^{2} &  & \text{final nucleons}
\end{array}
\end{equation}
$\vec{p}$ and $\overrightarrow{q_{\pi }}$ are, respectively, the initial
nucleon tri-momentum and the emitted pion tri-momentum. $\overrightarrow{%
q_{k}}\left( \overrightarrow{q_{u}}\right) $ is the relative tri-momentum of
the two intermediate(final) nucleons and $E_{\pi }=\sqrt{m_{\pi }^{2}+\left| 
\overrightarrow{q_{\pi }}\right| ^{2}}$ is the energy of the emitted pion.
All quantities are referred to the three-body center-of-mass frame of the $%
\pi NN$ final state.

\section{Treatment of the logarithmic singularity of the pion propagator (ISI)}
A partial wave decomposition involving two angles $\arccos x_{1}= \angle \left(
\overrightarrow{q_u}-\overrightarrow{q_{k}},\overrightarrow{q_{\pi }}\right)$ and 
$\arccos x_2 = \angle \left(\overrightarrow{q_u},\overrightarrow{q_{k}}\right)$
was implemented. In terms of 
$x_1$ and $x_2$ the pion propagator for the FSI amplitude given by Eq.(\ref{propagator})
reads
\begin{equation}
G_{\pi }=-\left( \frac{\alpha _{1}+\omega _{\pi }}{q_{\pi }I_{2}\left(
x_{2}\right) }\frac{1}{y_{1}-x_{1}}\right) \left( \frac{\alpha _{2}+\omega
_{\pi }}{q_{\pi }I_{2}\left( x_{2}\right) }\frac{1}{y_{2}-x_{1}}\right) 
\end{equation}
where $\alpha_1 = E-\frac{F_{1}-F_{2}}{2}-\omega _{2}-\frac{%
E_{\pi }}{2}$, $\alpha_2 = E+\frac{F_{1}-F_{2}}{2}-\omega _{1}
+\frac{E_{\pi }}{2}$ ($\alpha_1$ and $\alpha_2$ must 
be positive or zero) and $I_{2}\left( x_{2}\right) = \left| \overrightarrow{q_u}-
\overrightarrow{q_{k}} \right| =\sqrt{q_u^{2}-2 q_u q_{k}x_{2}+q_{k}^{2}}$. The 
roots of the denominator of $G_\pi$ are given by:
\begin{equation}
y_{i}=\frac{\alpha _{i}^{2}-\beta ^{2}}{q_{\pi }I_{2}\left( x_{2}\right) }%
\qquad \text{with }\beta ^{2}\equiv m_{\pi }^{2}+\frac{q_{\pi }^{2}}{4}%
+I_{2}\left( x_{2}\right) ^{2}\text{ and }i=1,2
\end{equation}

Once $G_{\pi }$ is in this form, we have to deal with its poles at 
$x_1=y_i$. We used the subtraction technique exposing the three-body 
logarithmic singularity:
\begin{eqnarray}
\int_{-1}^{1}\frac{f\left( x_{1},x_{2}\right) }{y_{i}-x_{1}}P_{L}\left(
x_{1}\right) dx_{1} &=&\mathbf{PV}\int_{-1}^{1}\frac{f\left(
x_{1},x_{2}\right) -f\left( y_{i},x_{2}\right) }{y_{1}-x_{1}}P_{L}\left(
x_{1}\right) +2f\left( y_{i},x_{2}\right) Q_{L}\left( y_{1}\right) - \nonumber \\
&&-i\pi P_{L}\left( y_{1}\right) f\left( y_{i},x_{2}\right) 
\end{eqnarray}
where $P_{L}$ and are the Legendre polynomials of order $L$ and $Q_L$ are the 
Legendre functions of the second kind of order $L$. These last functions exhibit logarithmic 
singularities which are given by the condition $y_i=\pm 1$. We determine its solution(s) 
analytically. Then, the integral over the moving logarithmic singularities is handled 
by a variable mesh. When the number of singularities (always between zero and two) is $n_s$,  the interval
of integration is divided into $2n_{s}+1$ regions with breaking points given by the found  singularities. 
In each one of the regions one considers a
Gaussian mesh. 
\pagebreak
\begin{center}
FIGURE CAPTIONS
\end{center}
\vspace{1cm}

Fig 1 --- Impulse, re-scattering and short-range processes which contribute to $\pi^0$ production. \\

Fig 2 --- Decomposition of the Feynman diagram in terms of six time-ordered diagrams for the final-state interaction. 
The pion(sigma) field is represented by a dashed(solid double) line. 
The nucleons are represented by solid lines. The DWBA amplitude may be identified to the first four
time-ordered diagrams ($a_1$ to $a_4$). The last two diagrams ($a_5$ to $a_6$) are usually called stretched boxes.\\

Fig 3 --- Decomposition of the Feynman diagram in terms of six time-ordered diagrams for the initial-state interaction.
The pion(sigma) field is represented by a dashed(solid double) line.
The nucleons are represented by solid lines. The DWBA amplitude corresponds to the first four time-ordered diagrams
($b_1$ to $b_4$) and the stretched boxes to the last two ($b_5$ to $b_6$).\\

Fig 4 --- Importance of the stretched boxes compared to the DWBA FSI amplitude $\mathcal{M}^{FSI}_{TOPT}$, 
as a function of the mass of the scalar particle for the final $NN$ interaction. 
The $\pi N $ amplitude is
contact re-scattering vertex (solid line) and the $\chi$Pt  (doted line).
The energy is taken at the pion production threshold. Absolute values of the amplitudes are considered.\\

Fig 5 --- Importance of the stretched boxes compared to the DWBA (FSI) for the cross section as a function of the mass 
of the scalar particle for the final $NN$ interaction. The $\pi N$ amplitude is the $\chi$Pt  
amplitude (left) and a contact re-scattering vertex (right).\\

Fig 6 --- The same as Fig.\ref{mfsi} but for the DWBA ISI amplitude $\mathcal{M}^{ISI}_{TOPT}$.\\

Fig 7 --- The same as Fig.\ref{sfsi} but for the ISI case.\\

Fig 8 --- Effect of the choices for the energy of the exchanged pion when the $\chi$Pt $\pi N$ re-scattering amplitude is considered, as a function of the laboratory kinetic energy. The nucleon-nucleon interaction is sigma exchange (panel a)) and 
the Bonn B potential (panel b)). The cross section curves shown correspond to FSI alone. 
The heavy full line is the reference calculation (Eq.(\ref{dwbafsi})). The dashed-dotted line is the {\it static} approximation. 
The dotted line is the {\it fixed 
kinematics} prescription and the dashed line is the {\it on-shell} prescription. The light full line and the light dotted line with 
bullets correspond respectively 
to the last two prescriptions, taken not only for the propagator but also for the $\pi N$ re-scattering  vertex.\\

Fig 9 --- The same as Fig.\ref{gfsi} but for the ISI amplitude.\\

Fig. 10 --- Imaginary part of the ISI amplitude $\mathcal{M}_{DWBA}$
as a function of the dimensionless pion momentum.
The exact result (full line) and {\it fixed kinematics} (dashed line), {\it on-shell}
(short-dashed line) and {\it static} (dotted line) approximations are shown.\\

Fig 11 --- Ratio between total approximated and exact cross section as a function of the dimensionless pion momentum. The {\it fixed kinematics} (full line), {\it on-shell}
(dashed line )and {\it static} (dotted line) cases are shown. \\

\pagebreak


\begin{thebibliography}{}
\bibitem{Park} Y. Park, F. Myhrer, J. R. Morones, T. Meissner and K. Kubodera, Phys. ReV. {\bf C70}, 1519 (1996).
\bibitem{Cohen} T. D. Cohen, J. L. Friar, G. A. Miller and U. van Kolck, Phys. Rev. {\bf C53}, 2661 (1996).
\bibitem{Sato} T. Sato, T.-S. H. Lee, F. Myhrer and K. Kubodera, Phys. Rev. {\bf C56}, 1246 (1997).

\bibitem{Riska} T.-S. Lee and D. O. Riska, Phys. Rev. Lett. {\bf 70}, 2237 (1993).

\bibitem{Oset} E. Hern\'andez and E. Oset, Phys. Lett. {\bf B350} 158 (1995).

\bibitem{Pena2} M. T. Pe\~na, S.A. Coon, J. Adam, Jr., A. Stadler,
Contributed to Bloomington 1999, Nuclear physics at storage rings,
AIP Conf.Proc.{\bf 512} 111 (2000). 

\bibitem{Hanh4} C. Hanhart, J. Haidenbauer, A. Reuber, C. Sch\"utz and J. Speth, Phys. Lett. {\bf B358}, 21 (1995). 


\bibitem{Kolck} U. van Kolck., G. A. Miller and D. O. Riska, Phys. Lett. {\bf B388}, 679 (1996).

\bibitem{Hanh0} C. Hanhart, J. Haidenbauer, O. Krehl and J. Speth, Phys. Lett {\bf B444} 25 (1998).

\bibitem{Pena} M. T. Pe\~na, D. O. Riska and A. Stadler, Phys. Rev. {\bf C60}, 045201 (1999).


\bibitem{Hanh00}
C. Hanhart, U. van Kolck, G.A. Miller, Phys.Rev.Lett.{\bf 85} (2000).

\bibitem{Epel} E. Epelbaum, A. Nogga, W. Glockle, H. Kamada, U.G. Meissner and H. Witala, Eur. Phys. J. {\bf A15}, 543 (2002).

\bibitem{NewHanh} C. Hanhart and N. Kaiser, Phys. Rev. {\bf C66}, 054005 (2002).

\bibitem{Hanh1} C. Hanhart, G. A. Miller, F. Myhrer, T. Sato and U. van Kolck, Phys. Rev. {\bf C63}, 044002 (2001).

\bibitem{Hanh2} A. Motzke, Ch. Elster and C. Hanhart, Phys. Rev. {\bf C66}, 054002 (2002).

\end{thebibliography}
\end{document}